\documentclass[aps,pra,superscriptaddress]{revtex4-2}

\usepackage{amsmath,amsfonts,amssymb}
\usepackage{xcolor,graphicx}
\usepackage{braket}
\usepackage{float}
\usepackage[pdftex]{hyperref}
\hypersetup{
	colorlinks = true,
	linkcolor = blue,
	anchorcolor = blue,
	citecolor = red,
	filecolor = blue,
	urlcolor = blue}
\usepackage{subfig}
\usepackage{overpic}

\usepackage{listings}
\usepackage{xcolor}
\usepackage[utf8]{inputenc}
\usepackage{float}

\lstdefinestyle{python}{
    language=Python,
    backgroundcolor=\color{white},   
    basicstyle=\footnotesize\ttfamily, 
    breaklines=true,                  
    captionpos=b,                     
    numbers=left,                     
    numberstyle=\tiny\color{gray},   
    keywordstyle=\color{blue},        
    commentstyle=\color{green!80!black}, 
    stringstyle=\color{red},          
}

\begin{document}

\title{Numerical solution of the Lindblad master equation using the Runge-Kutta method implemented in Python}

\author{L. Hernández-Sánchez}
\email[e-mail: ]{leonardi1469@gmail.com}
\affiliation{Instituto Nacional de Astrofísica Óptica y Electrónica, Calle Luis Enrique Erro No. 1\\ Santa María Tonantzintla, Puebla, 72840, Mexico}
\author{I. A. Bocanegra-Garay}
\affiliation{Departamento de F\'isica Te\'orica, At\'omica y \'Optica and  Laboratory for Disruptive Interdisciplinary Science (LaDIS), Universidad de Valladolid, 47011 Valladolid, Spain}
\author{A. Flores-Rosas}
\affiliation{Facultad de Ciencias en F\'isica y Matem\'aticas, Universidad Aut\'onoma de Chiapas, Carretera Emiliano Zapata, Km. 8, Rancho San Francisco, 29050 Tuxtla Guti\'errez, Chiapas, Mexico}
\author{I. Ramos-Prieto}
\affiliation{Instituto Nacional de Astrofísica Óptica y Electrónica, Calle Luis Enrique Erro No. 1\\ Santa María Tonantzintla, Puebla, 72840, Mexico}
\author{F. Soto-Eguibar}
\affiliation{Instituto Nacional de Astrofísica Óptica y Electrónica, Calle Luis Enrique Erro No. 1\\ Santa María Tonantzintla, Puebla, 72840, Mexico}
\author{H.M. Moya-Cessa}
\affiliation{Instituto Nacional de Astrofísica Óptica y Electrónica, Calle Luis Enrique Erro No. 1\\ Santa María Tonantzintla, Puebla, 72840, Mexico}

\date{\today}

\begin{abstract}
The dynamics of open quantum systems is governed by the Lindblad master equation, which provides a consistent framework for incorporating environmental effects into the evolution of the system. Since exact solutions are rarely available, numerical methods become essential tools for analyzing such systems. This article presents a step-by-step implementation of the fourth-order Runge-Kutta method in Python to solve the Lindblad equation for a single quantized field mode subject to decay. A coherent state is used as the initial condition, and the time evolution of the average photon number is investigated. The proposed methodology enables transparent and customizable simulations of dissipative quantum dynamics, emphasizing a pedagogical approach that helps readers understand the numerical structure without relying on external libraries such as QuTiP. This standalone implementation offers full control over each integration step, making it particularly suitable for educational contexts and for exploring non-standard dynamics or introducing custom modifications to the Liouvillian.\\
\textbf{Palabras clave}: Ecuación de Schrödinger, interacción radiación-materia, método de Runge-Kutta, modelo de Jaynes-Cummings, Python.
\end{abstract}
\maketitle

\section{Introduction}\label{Introducción}
The study of open quantum systems is essential to understand how real systems interact with their environment. Unlike closed systems, whose evolution is purely coherent and reversible according to the Schrödinger equation, open systems undergo processes of dissipation and decoherence, which degrade quantum coherence due to coupling with an external environment. These effects are unavoidable in practice and are fundamental in areas such as quantum optics, quantum computing, quantum information, and spectroscopy~\cite{Breuer_Book, Carmichael_Book, Gerry_Book}.

The most general mathematical tool to describe the dynamics of such systems is the Lindblad master equation, which consistently models the time evolution of the density matrix by incorporating both coherent dynamics and irreversible effects induced by the environment~\cite{Arévalo_1995, Manzano_2020}. Although this equation has a compact and elegant form, exact solutions are only possible in particular cases, due to the complexity of simultaneously treating both unitary and dissipative terms~\cite{Hernandez_2023, Hernandez_2024, Hernandez_2025}.

In this context, numerical methods have become indispensable tools for studying the evolution of open systems. Among these, the fourth-order Runge-Kutta (RK4) method stands out for its balance between accuracy, simplicity, and ease of implementation to solve ordinary differential equations, making it an excellent option for approximating solutions of the Lindblad equation in increasingly complex systems~\cite{Butcher_book, Press_book}.

In previous studies focused on closed systems, the effectiveness of the RK4 method in simulating the dynamics of radiation-matter interaction models governed by the Schrödinger equation has been demonstrated~\cite{Hernandez_book}. In the present work, this strategy is extended to open quantum systems by presenting a detailed implementation of the RK4 method to numerically solve the Lindblad master equation.

As an illustrative example, the case of a single quantized field mode subject to decay is analyzed, evolving from an initial coherent state. The implementation of this simulation in Python is presented, including the computation of the average photon number as a physical observable and its comparison with the exact analytical solution, thereby validating the accuracy of the proposed approach.

Beyond its numerical utility, the methodology is developed with a pedagogical emphasis, enabling readers to grasp the structure of the algorithm and the role of each term in the evolution equation without relying on specialized packages such as QuTiP. This educational perspective also provides the flexibility needed to extend the method to more general or nonstandard dissipative scenarios, such as time-dependent Hamiltonians or state-dependent dissipation.

The main objective of this work is to provide a clear and accessible guide to simulate the dynamics of open quantum systems using a basic computational approach, promoting the understanding of these tools in both educational and research contexts.

The structure of the article is as follows. Section~\ref{Modelo} presents the physical model and its analytical solution. Section~\ref{Metodología} describes step by step the implementation of the RK4 method to solve the Lindblad equation. In Section~\ref{Resultados}, numerical results are compared with the analytical solution. Finally, Section~\ref{Conclusiones} presents the general conclusions of the work.

\section{The physical model and its analytical solution}\label{Modelo}
This section describes the physical system analyzed throughout the article: a single mode of a quantized electromagnetic field confined in a lossy optical cavity. This system can be interpreted as a quantum harmonic oscillator that loses energy due to its interaction with the environment~\cite{Louisell_1990}.

\begin{figure}[H]
\centering
\includegraphics[width=0.6\linewidth]{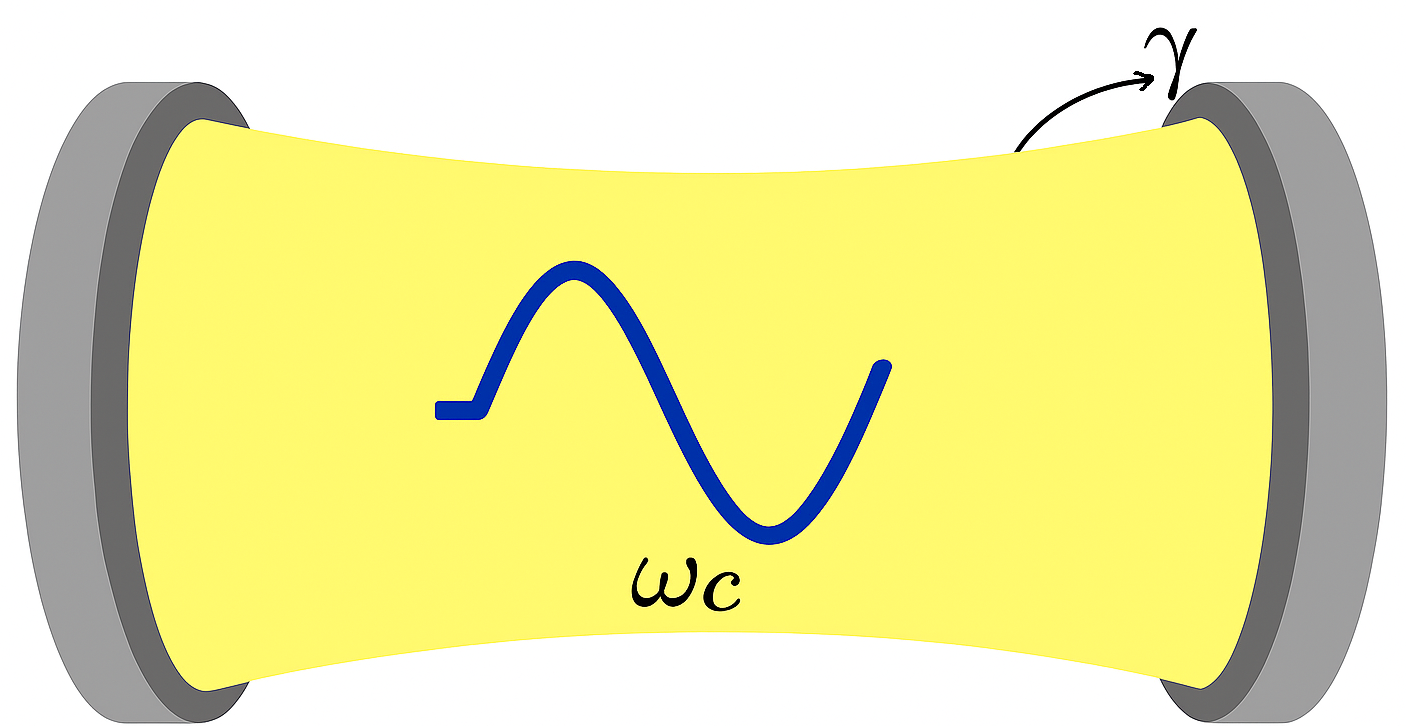}
\caption{Schematic representation of a single quantized field mode with frequency $\omega_c$, confined in an optical cavity with losses characterized by decay rate $\gamma$.}
\label{Cavidad}
\end{figure}

Figure~\ref{Cavidad} shows a schematic representation of the considered system. A single field mode with frequency $\omega_c$ is confined between two semitransparent mirrors. The interaction with the environment causes photons to leak out of the cavity, modeled through a decay process characterized by the rate $\gamma$.

The time evolution of the quantum state of the system, described by the density matrix $\hat{\rho}(t)$, is governed by the Lindblad master equation. For this setup, the free Hamiltonian is:
\begin{equation}
\hat{H} = \omega_c \hat{a}^\dagger \hat{a},
\end{equation}
where $\hat{a}$ and $\hat{a}^\dagger$ are the annihilation and creation operators of the field mode, and $\omega_c$ is its frequency.

The coupling to the environment is modeled using a single collapse operator:
\begin{equation}
\hat{C} = \sqrt{\gamma} \hat{a},
\end{equation}
where $\gamma$ represents the photon loss rate.

The Lindblad master equation for this system takes the form~\cite{Breuer_Book, Carmichael_Book, Gerry_Book}:
\begin{equation}\label{Lidblad}
\frac{d\hat{\rho}}{dt} = -i[\hat{H}, \hat{\rho}] + \gamma \left( \hat{a} \hat{\rho} \hat{a}^\dagger - \frac{1}{2} \left\{ \hat{a}^\dagger \hat{a}, \hat{\rho} \right\} \right).
\end{equation}
Each term in this equation has a specific role:
\begin{itemize}
\item The commutator $-i[\hat{H}, \hat{\rho}]$ represents the unitary evolution of the system, analogous to the Schrödinger equation but applied to the density matrix.
\item The term $\hat{a} \hat{\rho} \hat{a}^\dagger$ describes quantum jump processes, in which the system emits a photon.
\item The anticommutator terms $-\frac{1}{2} \{ \hat{a}^\dagger \hat{a}, \hat{\rho} \}$ account for the gradual loss of coherence due to environmental coupling.
\end{itemize}

In this work, the system is initialized in a coherent state $|\alpha\rangle$, which is an eigenstate of the annihilation operator: $\hat{a} |\alpha\rangle = \alpha |\alpha\rangle$. This type of state describes a quantum field whose statistical properties resemble those of a classical wave~\cite{Gerry_Book}.

Although the Lindblad equation generally does not admit analytical solutions, this particular case, a single-field mode subject to decay, does have a known solution. As shown in Chapter~5 of~\cite{Moya_Book}, a coherent state remains coherent throughout the evolution, with an amplitude that decays exponentially in time. The density matrix at time $t$ is:
\begin{equation}\label{Sol}
\hat{\rho}(t) = |\alpha(t)\rangle \langle \alpha(t)|,
\end{equation}
where the amplitude evolves as:
\begin{equation}
\alpha(t) = \alpha \, e^{-\frac{\gamma}{2} t} \, e^{-i \omega_c t}.
\end{equation}

A physically relevant observable characterizing the system dynamics is the average number of photons. In the density matrix formalism, it is computed as:
\begin{equation}
\langle \hat{n} \rangle (t) = \mathrm{Tr} \left[ \hat{a}^\dagger \hat{a} \, \hat{\rho}(t) \right].
\end{equation}
Since $\hat{\rho}(t)$ remains a pure coherent state and $\hat{a} |\alpha(t)\rangle = \alpha(t) |\alpha(t)\rangle$, it follows that:
\begin{equation}\label{Valor de n}
\langle \hat{n} \rangle (t) = \langle \alpha(t)| \hat{a}^\dagger \hat{a} | \alpha(t) \rangle = |\alpha(t)|^2 = |\alpha|^2 e^{-\gamma t}.
\end{equation}
This expression shows that the average photon number decays exponentially at a rate $\gamma$, reflecting the energy loss of the field due to its coupling to the environment.

For simplicity, this work considers a zero temperature environment, which allows us to focus on numerical implementation without additional thermal excitation effects. In this regime, the system evolves towards a stationary state as $t \to \infty$, where the density matrix tends to $\hat{\rho}_{\text{ss}} = |0\rangle\langle 0|$. This vacuum state represents the complete decay of the field energy, and the exact steady-state value of the average photon number is $\langle \hat{n} \rangle_{\text{ss}} = 0$.

However, the model can be naturally extended to finite-temperature environments by incorporating additional dissipative terms that depend on the mean thermal photon number $\bar{n}$. In such cases, the system asymptotically approaches a thermal mixed state instead of the vacuum, and the steady-state average photon number becomes $\langle \hat{n} \rangle_{\text{ss}} = \bar{n}$. This thermalization behavior, along with the uniqueness and structure of the steady state, can be directly understood from the spectral properties of the Liouvillian superoperator~\cite{Arévalo_1995}.

\section{Methodology}\label{Metodología}
This section provides a detailed description of the application of the RK4 method to solve the Lindblad master equation~\eqref{Lidblad} for the case of a single quantized field mode subject to decay. This approach not only offers high accuracy and ease of implementation, but can also be naturally extended to more complex models, such as systems with multiple interacting modes, coupling to thermal reservoirs\textcolor{red}{,} or several dissipation channels, situations in which analytical solutions are generally unavailable.

\subsection{Problem formulation}
As discussed previously, the dynamics of the system is fully determined by the time evolution of its density matrix $\hat{\rho}(t)$, which is obtained by solving the Lindblad master equation~\eqref{Lidblad}. In its general form, this equation can be written as:
\begin{equation}\label{Lidblad_2}
\frac{d\hat{\rho}}{dt} = \mathcal{L}[\hat{\rho}],
\end{equation}
where $\mathcal{L}$ denotes the \textit{Lindblad superoperator}, which encapsulates both the coherent dynamics of the system and the dissipative effects arising from its interaction with the environment.

Formally, the solution to this equation can be expressed as:
\begin{equation}\label{Solution}
\hat{\rho}(t) = e^{\mathcal{L} t} \hat{\rho}(0),
\end{equation}
where $\hat{\rho}(0) = |\psi(0)\rangle \langle \psi(0)|$ represents the initial condition of the system.

The main objective of this section is to develop, step by step, a methodology based on the RK4 algorithm and to implement it in the Python programming language\textcolor{red}{,} in order to approximate the solution of the equation~\eqref{Lidblad_2}. As an illustrative example, the model of a single-field mode undergoing decay is considered. The analytical solution previously presented allows for direct validation of the numerical results.

It is important to note that this methodology is not limited to the specific case analyzed here. It can be readily generalized to more complex quantum systems, including multiple field modes, interaction with atoms, coupling to thermal environments, or arbitrary initial conditions.

\subsection{Fourth-order Runge-Kutta method adapted to matrices}

The RK4 method is a widely used numerical technique for solving ordinary differential equations (ODE) of the form~\cite{Butcher_book}:
\begin{equation}
\frac{dy}{dt} = f(t, y), \quad y(0) = y_0.
\end{equation}

The core idea of RK4 is to approximate the solution using a weighted combination of several evaluations of the function $f(t, y)$ within each finite time increment $\Delta t$, which represents the time step used in the numerical integration. The update scheme is expressed as:
\begin{align}
t_{i+1} &= t_i + \Delta t,\\
k_1 &= f(t_i, y_i), \\
k_2 &= f\left(t_i + \frac{\Delta t}{2}, y_i + \frac{\Delta t}{2} k_1\right), \\
k_3 &= f\left(t_i + \frac{\Delta t}{2}, y_i + \frac{\Delta t}{2} k_2\right), \\
k_4 &= f\left(t_i + \Delta t, y_i + \Delta t\, k_3\right), \\
y_{i+1} &= y_i + \frac{\Delta t}{6} \left(k_1 + 2k_2 + 2k_3 + k_4\right).
\end{align}

Its main advantage lies in its fourth-order accuracy, which means that the local error per time step is on the order of $\mathcal{O}(\Delta t^5)$, while the total cumulative error is $\mathcal{O}(\Delta t^4)$. Although this method requires four evaluations of the function $f$ per time step, RK4 offers an excellent balance between accuracy and computational efficiency.

In the context of the Lindblad master equation, the function $f(t, y)$ is replaced by the action of the Lindblad superoperator $\mathcal{L}$ on the density matrix:
\begin{equation}
f(t, \hat{\rho}) = \mathcal{L}[\hat{\rho}],
\end{equation}
where $\hat{\rho}(t)$ is a Hermitian matrix that describes the quantum state of the system. The RK4 scheme can be applied without formal modification, although the operations must now be interpreted as acting on complex matrices rather than on scalars or vectors. In this case, the variable $y$ is identified with the density matrix $\hat{\rho}$, and the quantities $k_1$, $k_2$, $k_3$, and $k_4$ are also matrices of the same dimension, calculated iteratively by the action of the superoperator $\mathcal{L}$.

One of the principal advantages of this method is its ease of implementation in programming languages such as Python, making it an attractive tool for studying the dynamics of open quantum systems. Although its computational cost may be higher than that of simpler algorithms, such as the Euler method, its accuracy and numerical stability make it especially useful for simulating systems without analytical solutions where reliable numerical integration is required.

It is worth noting that this implementation employs a fixed time step $\Delta t$, which helps to clearly illustrate the structure of the algorithm and ensures consistency across different numerical runs. This choice aligns with the pedagogical purpose of this work, which is to offer a transparent and accessible introduction to the numerical integration of the Lindblad master equation.

However, in scenarios where the system exhibits fast transients or widely separated time scales—such as strong dissipation regimes or complex interactions—using a fixed step size may lead to either poor accuracy or inefficient use of computational resources. In such cases, it is advisable to employ adaptive step-size techniques that dynamically adjust $\Delta t$ based on an estimate of the local truncation error.

A well-established solution is the Runge-Kutta-Fehlberg (RKF) method, which combines two embedded Runge-Kutta formulas of different orders (typically fourth and fifth) to control the error and adapt the time step accordingly~\cite{Fehlberg_1969, Hammachukiattikul_2021, Nwankwo_2021}. The methodology presented in this work can be naturally extended to incorporate such schemes, offering higher numerical efficiency without altering the underlying structure of the integration algorithm.

\subsection{Implementation in Python}
This subsection presents the implementation of the RK4 method in Python to numerically solve the Lindblad master equation~\eqref{Lidblad_2}, applied to the model of a single quantized field mode subject to decay. The code developed here applies the RK4 iterative scheme directly in the matrix context, meaning that the variables involved are treated as operators (i.e., dense complex matrices) rather than scalar functions. The time evolution is computed on a truncated Fock basis of finite dimension, which is appropriate for representing initial coherent states.

The following code illustrates the implementation of this methodology step by step.
\begin{enumerate}
\item \textbf{Library importation}

The first step in the implementation consists of importing the fundamental libraries that allow for performing numerical operations, manipulating matrix structures, and generating graphical representations of the results. These tools are essential for defining the operators of the quantum system, computing the time evolution using the RK4 method, and visualizing the dynamic behavior of the system.

\begin{lstlisting}[style=python, caption={}]
import numpy as np
import matplotlib.pyplot as plt
import math
from scipy.linalg import null_space
\end{lstlisting}

\begin{itemize}
    \item \texttt{numpy}: This is the standard Python library for numerical calculations with multidimensional arrays. In this work, it is used to define the system operators, manipulate the density matrix, and carry out matrix multiplications and summations required for the time evolution.
    
    \item \texttt{matplotlib.pyplot}: This library makes it possible to create two-dimensional plots. In the context of quantum simulation, it is used to visualize the time evolution of physical observables, thereby facilitating the interpretation of the results.
    
    \item \texttt{math}: This library provides standard mathematical functions. In particular, it is used to compute exact integer factorials, which are necessary for constructing the initial coherent state as a superposition of Fock states.
    
    \item \texttt{scipy.linalg.null\_space}: This function is used to compute the steady state of the system by finding the null space of the Liouvillian superoperator. It allows for direct numerical access to the long-time limit of the dynamics, complementing the RK4 simulation.
\end{itemize}

\item \textbf{Definition of the system parameters}

In this step, the fundamental parameters of the model are defined. These parameters describe the physical properties and initial conditions of the quantum system under study.

\begin{lstlisting}[style=python, caption={}]
N       = 30        
omega_c = 0.9       
gamma   = 0.3       
alpha   = 3.0       
t_f     = 20        
dt      = 0.01      
t_list  = np.arange(0, t_f, dt) 
\end{lstlisting}

\begin{itemize}
    \item \texttt{N}: Specifies the dimension of the truncated Hilbert space, which corresponds to the maximum number of Fock states used to represent the field. 
    
    \item \texttt{omega\_c}: Sets the angular frequency of the single quantized mode of the electromagnetic field confined in the optical cavity.
    
    \item \texttt{gamma}: Represents the decay rate of the cavity, which models the photon loss due to coupling with the external environment. 
    
    \item \texttt{alpha}: Determines the complex amplitude of the initial coherent state $\ket{\alpha}$ used to initialize the system. 
    
    \item \texttt{t\_f}: Specifies the total duration of the simulation. The evolution of the system will be computed from $t = 0$ to $t = t_f$.
    
    \item \texttt{dt}: Defines the size of each discrete time step used by the Runge-Kutta integration algorithm. Smaller values improve accuracy but require more computational resources.
    
    \item \texttt{t\_list}: Creates a NumPy array of equally spaced time points using the function \texttt{np.arange}. This array defines the specific time instants at which the observables of the system will be evaluated during the simulation.
\end{itemize}

\item \textbf{Definition of operators and system Hamiltonian}

This block defines the fundamental operators that describe the quantum dynamics of a single field mode confined in a cavity. These operators include the annihilation operator $\hat{a}$, the creation operator $\hat{a}^\dagger$, and the number operator $\hat{n} = \hat{a}^\dagger \hat{a}$. From these, the free Hamiltonian of the system is constructed, as well as the collapse operator that accounts for the dissipative interaction of the system with the environment.

\begin{lstlisting}[style=python, caption={}]
def destroy(N):
    return np.diag(np.sqrt(np.arange(1, N)), 1)
a = destroy(N)
ad = a.conj().T
n_op = ad @ a
H = omega_c * n_op
c = np.sqrt(gamma) * a
\end{lstlisting}

\begin{itemize}
    \item \texttt{destroy(N)}: This function returns the annihilation operator $\hat{a}$ represented as an $N \times N$ matrix in the truncated Fock basis.

    \item \texttt{a}: Represents the annihilation operator $\hat{a}$, which lowers the photon number of the field mode by one.

    \item \texttt{ad}: Represents the creation operator $\hat{a}^\dagger$, obtained as the complex conjugate transpose of the annihilation operator. It increases the photon number by one when acting on a Fock state.

    \item \texttt{n\_op}: Defines the number operator $\hat{n} = \hat{a}^\dagger \hat{a}$ using the matrix product \texttt{ad @ a}. In Python, the symbol \texttt{@} is used to perform matrix multiplication, unlike \texttt{*}, which denotes element-wise multiplication.

    \item \texttt{H}: Constructs the Hamiltonian $\hat{H} = \omega_c \hat{n}$, which describes the unitary evolution of the quantized field mode with frequency $\omega_c$ in the absence of interaction with the environment.

    \item \texttt{c}: Defines the collapse operator $\hat{C} = \sqrt{\gamma} \hat{a}$, which characterizes the photon loss process due to coupling with the external environment.
\end{itemize}

\item \textbf{Definition of the Lindblad master equation}

In this step, the Lindblad master equation is formulated to describe the time evolution of the density matrix $\hat{\rho}(t)$ in an open quantum system. This equation incorporates both the coherent dynamics governed by the system Hamiltonian and the dissipative effects arising from its interaction with the environment.

\begin{lstlisting}[style=python, caption={}]
def lindblad_rhs(rho, H, c_ops):
    commutator = -1j * (H @ rho - rho @ H)
    dissipator = np.zeros_like(rho, dtype=complex)
    for c in c_ops:
        dissipator += c @ rho @ c.conj().T - 0.5 * (c.conj().T @ c @ rho + rho @ c.conj().T @ c)
    return commutator + dissipator
\end{lstlisting}

\begin{itemize}
\item \texttt{lindblad\_rhs(rho, H, c\_ops)}: This function computes the right-hand side of the Lindblad master equation.
\item \texttt{rho}: Represents the density matrix of the system at a given time.
\item \texttt{H}: The system Hamiltonian, which generates the unitary part of the evolution.
\item \texttt{c\_ops}: A list of collapse operators $\hat{C}_n$ that model dissipation channels. In this particular model, only one collapse operator is used: $\hat{C} = \sqrt{\gamma} \hat{a}$.
\item \texttt{commutator}: Computes the coherent term $-i[\hat{H}, \hat{\rho}]$, corresponding to unitary evolution.
\item \texttt{dissipator}: Computes the dissipative part of the Lindblad equation, which accounts for the non-unitary evolution due to the interaction with the environment.
\end{itemize}

\item \textbf{Implementation of the RK4 method}

In this step, the RK4 method is applied to numerically solve the Lindblad master equation. This technique approximates the time evolution of the density matrix $\hat{\rho}(t)$ by dividing the time domain into discrete steps and computing the solution iteratively.

\begin{lstlisting}[style=python, caption={}]
def rk4_step(rho, H, c_ops, dt):
    k1 = lindblad_rhs(rho, H, c_ops)
    k2 = lindblad_rhs(rho + dt/2 * k1, H, c_ops)
    k3 = lindblad_rhs(rho + dt/2 * k2, H, c_ops)
    k4 = lindblad_rhs(rho + dt * k3, H, c_ops)
    return rho + (dt/6) * (k1 + 2*k2 + 2*k3 + k4)
\end{lstlisting}

\begin{itemize}
\item \texttt{rk4\_step}: This function performs a single integration step using the RK4 method applied to the Lindblad equation. It takes as input the current density matrix \texttt{rho}, the system Hamiltonian \texttt{H}, the list of collapse operators \texttt{c\_ops}, and the time step size \texttt{dt}.
\item \texttt{k1, k2, k3, k4}: These are the four intermediate evaluations required by the RK4 scheme to estimate the derivative of the density matrix. Each one corresponds to a different point in time and uses a different approximation of the state.
\item Each \texttt{k} is computed by calling the \texttt{lindblad\_rhs} function, which returns the value of the time derivative at the corresponding state.
\item The weighted combination of the four terms (\texttt{k1 + 2*k2 + 2*k3 + k4}) produces a fourth-order accurate estimate of the state at the next time step.
\item The final result is the updated density matrix at the next time step, computed with high precision and numerical stability.
\end{itemize}

\item \textbf{Definition of the initial state of the system}

In this step, the initial condition of the system is established. As the initial state, a coherent state is selected. Instead of representing this state as a pure state vector, its corresponding density matrix $\hat{\rho}_0 = \ket{\alpha}\bra{\alpha}$ is used, which is necessary for solving the Lindblad master equation.

\begin{lstlisting}[style=python, caption={}]
def coherent_dm(N, alpha):
    vec = np.array([alpha**m / math.sqrt(math.factorial(m)) for m in range(N)], dtype=complex)
    vec *= np.exp(-0.5 * abs(alpha)**2)
    return np.outer(vec, vec.conj())
rho_0 = coherent_dm(N, alpha)
\end{lstlisting}
\begin{itemize}
\item \texttt{coherent\_dm(N, alpha)}: This function constructs the density matrix of the coherent state $\ket{\alpha}$ in a Fock basis truncated to dimension $N$.
\item \texttt{np.outer(vec, vec.conj())}: Computes the outer product between a complex vector $\ket{\psi}$ and its conjugate transpose $\bra{\psi}$, resulting in a pure state density matrix. In this context, the use of \texttt{np.outer} ensures the construction of a square matrix of size $N \times N$ that accurately represents the initial field state.
\item \texttt{rho\_0}: Denotes the initial density matrix of the system, which is used as the input for the evolution governed by the Lindblad master equation.
\end{itemize}

\item \textbf{Time evolution of the density matrix and calculation of observables}

Once the time evolution of the density matrix $\hat{\rho}(t)$ has been obtained, it is possible to calculate any physical observable of the system. In this work, the mean photon number $\langle \hat{n}(t) \rangle$ is evaluated, which provides a direct measure of the energy stored in the field mode over time.

\begin{lstlisting}[style=python, caption={}]
n_expect = []
for t in t_list:
    n_expect.append(np.real(np.trace(n_op @ rho_0)))
    rho_0 = rk4_step(rho_0, H, [c], dt)
\end{lstlisting}
\begin{itemize}
    \item \texttt{n\_expect}: A list that stores the mean photon number $\langle \hat{n}(t) \rangle$ at each time step.
    \item \texttt{rho\_0}: Represents the density matrix that is updated at each time step using the function \texttt{rk4\_step}, previously defined. In each iteration, it is replaced by the matrix at the next time step.
    \item \texttt{np.trace(n\_op @ rho\_0)}: Computes the expectation value of the number operator in the current state using the formula $\langle \hat{n} \rangle = \mathrm{Tr}(\hat{n} \hat{\rho})$. The result is converted to its real part using \texttt{np.real()} to eliminate possible small imaginary components arising from numerical rounding.
\end{itemize}

\end{enumerate}

\subsection{Method validation}

To verify the precision of the numerical implementation of the RK4 method applied to the Lindblad master equation, we compare the results for a physical observable with its corresponding analytical solution. Specifically, we analyze the time evolution of the average photon number $\langle \hat{n}(t) \rangle$ in a single-mode quantum field initially prepared in a coherent state $\ket{\alpha}$ and subject to losses characterized by a decay rate $\gamma$. The exact analytical solution for this case is given by equation \eqref{Valor de n}.

To carry out the validation, we use the analytical expression for the mean photon number $\langle \hat{n}(t) \rangle$, derived at the end of Section~\ref{Modelo}, where both the transient and long-time behaviors were discussed. This solution provides a direct benchmark for the numerical results obtained with the RK4 method and serves as a reference for the asymptotic behavior of the system at zero temperature.

\begin{lstlisting}[style=python, caption={}]
n_analytic = np.abs(alpha)**2 * np.exp(-gamma * t_list)
n_ss_analytic = 0.0
\end{lstlisting}

To confirm that the numerical method correctly captures the steady-state behavior of the open quantum system, the Liouvillian superoperator is constructed explicitly in matrix form and its null space is computed. The steady state corresponds to the eigenvector associated with the zero eigenvalue of this operator.

\begin{lstlisting}[style=python, caption={}]
I = np.eye(N)
L = -1j * (np.kron(I, H) - np.kron(H.T, I))
L += np.kron(c, c.conj()) - 0.5 * (np.kron(I, c.conj().T @ c) + np.kron((c.conj().T @ c).T, I))
L = L.reshape(N*N, N*N)

nullvec = null_space(L)
if nullvec.shape[1] == 0:
    raise ValueError("No steady state found (null space is empty).")

rho_ss_vec = nullvec[:, 0]
rho_ss = rho_ss_vec.reshape((N, N))
rho_ss = rho_ss / np.trace(rho_ss)

n_ss_liouvillian = np.real(np.trace(n_op @ rho_ss))
\end{lstlisting}

\begin{itemize}
    \item \texttt{n\_analytic}: Computes the analytical expression for the average photon number in a decaying coherent state.
    \item \texttt{n\_ss\_analytic}: Stores the expected steady-state value $\langle \hat{n} \rangle$ at zero temperature, which corresponds to the vacuum.
    \item \texttt{L}: Constructs the Liouvillian superoperator in vectorized form using Kronecker products to represent the Lindblad master equation as a matrix equation.
    \item \texttt{null\_space(L)}: Computes the null space of the Liouvillian matrix. A non-trivial eigenvector (if found) corresponds to the steady state $\rho_{\mathrm{ss}}$.
    \item \texttt{rho\_ss}: The steady-state density matrix, reshaped and normalized.
    \item \texttt{n\_ss\_liouvillian}: Expectation value of the number operator evaluated in the steady state obtained from the Liouvillian formalism.
\end{itemize}

The final comparison of results is shown graphically:

\begin{lstlisting}[style=python, caption={}]
plt.figure(figsize=(8, 5))
plt.plot(t_list, n_expect, label="RK4 (numerical)", linewidth=2.5, color='orange')
plt.plot(t_list, n_analytic, '--', label="Analytical solution", linewidth=1.5, color='blue')
plt.axhline(y=n_ss_liouvillian, color='skyblue', linestyle='-', linewidth=2.5,
            label="Steady state (Liouvillian)")
plt.axhline(y=n_ss_analytic, color='green', linestyle=':', linewidth=1.5,
            label="Analytical steady state")
plt.xlabel(r"Time $t$", fontsize=14)
plt.ylabel(r"$\langle \hat{n}(t) \rangle$", fontsize=14)
plt.title("Mean photon number: evolution and steady state", fontsize=14)
plt.legend()
plt.grid(True)
plt.tight_layout()
plt.savefig("Average photon number.pdf", format="pdf")
plt.show()
\end{lstlisting}

\begin{itemize}
    \item \texttt{plt.plot(t\_list, n\_expect, ...)}: Plots the numerical values of the average photon number obtained using the RK4 method (orange line).
    \item \texttt{plt.plot(t\_list, n\_analytic, ...)}: Plots the exact analytical solution as a dashed blue line.
    \item \texttt{plt.axhline(..., n\_ss\_liouvillian, ...)}: Draws a horizontal line showing the steady-state value from the Liouvillian approach (sky blue).
    \item \texttt{plt.axhline(..., n\_ss\_analytic, ...)}: Draws a horizontal line indicating the analytically expected steady-state value (green dotted line).
    \item The graph offers a visual validation of the method: the numerical and analytical curves overlap during the transient regime and converge to the same steady-state value.
\end{itemize}

\noindent The resulting graph is shown in Figure~\ref{Valor esp de n}:
\begin{figure}[H]
    \centering
    \includegraphics[width=0.6\linewidth]{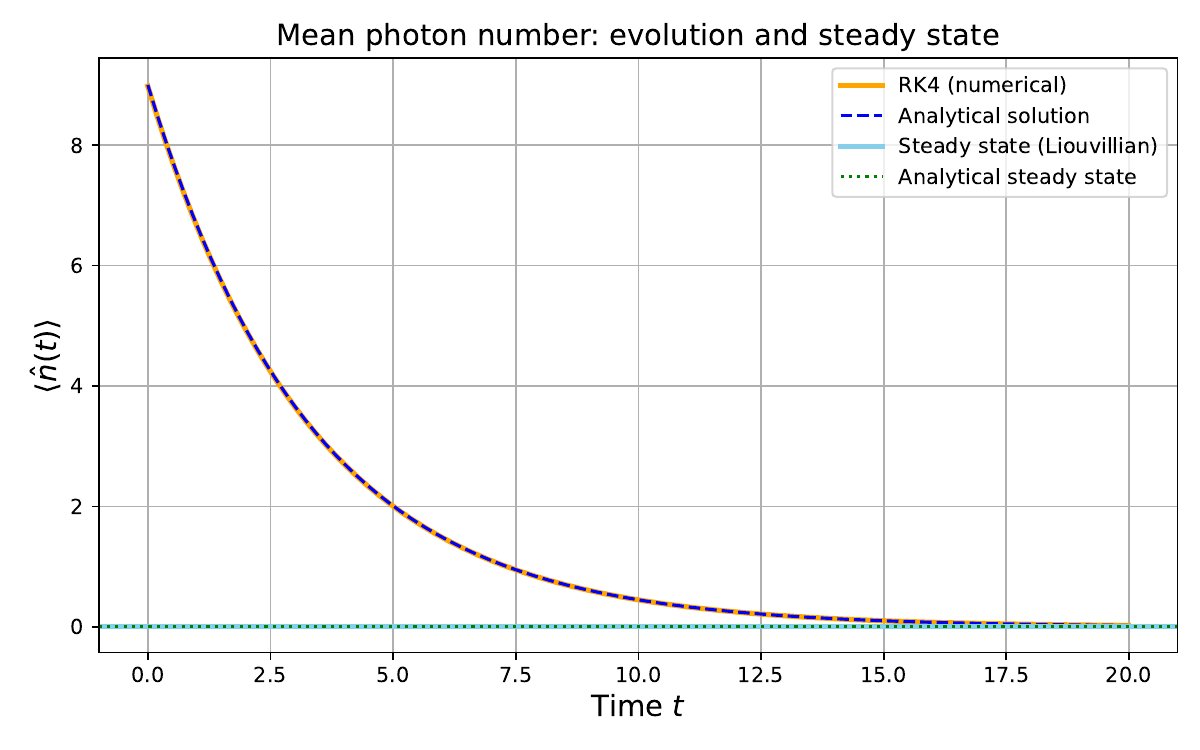}
    \caption{Time evolution of the average photon number $\langle \hat{n}(t) \rangle$ for a single-mode decaying quantum field. The solid orange line corresponds to the numerical solution obtained using the RK4 method, the dashed blue line represents the exact analytical result, and the horizontal lines indicate the steady-state predictions from both approaches.}
    \label{Valor esp de n}
\end{figure}

\section{Results and discussion}\label{Resultados}
The numerical implementation of the RK4 method to solve the Lindblad master equation exhibits excellent agreement with the exact analytical solution for the average photon number in a decaying coherent field mode. This correspondence, shown in Figure~\ref{Valor esp de n}, validates both the precision and the numerical stability of the proposed scheme.

The numerical results accurately reproduce the exponential decay law predicted by theory for the time evolution of the average number of photons. The trajectory obtained using RK4 remains close to the analytical solution throughout the simulated interval, without displaying significant cumulative deviations.

In addition to verifying the transient behavior, the implementation also recovers the expected steady state by computing the null space of the Liouvillian superoperator. This numerical extraction of the stationary state confirms that the method captures both short- and long-term dynamics. The ability to access the steady state in matrix form further enriches the applicability of the code for the analysis of equilibrium and spectral studies of open quantum systems.

Although existing high-level libraries such as \texttt{QuTiP} provide efficient and user-friendly solvers for quantum master equations, they encapsulate the numerical procedure behind abstraction layers. In contrast, the RK4 implementation developed here explicitly reveals each step of the time evolution, making it particularly valuable for educational purposes. This clarity helps learners to understand how dissipation enters the dynamics and how the Lindblad structure is numerically realized.

By implementing the algorithm from scratch, users gain full control over the numerical evolution, enabling them to explore modified Hamiltonians, time-dependent interactions, or nonstandard dissipative terms—scenarios that may not be directly accessible with prebuilt solvers.

Thus, the proposed approach is not intended to complement libraries like \texttt{QuTiP}, but rather to reveal and demystify the internal workings that such libraries typically conceal. This makes it especially suitable for pedagogical contexts and for researchers looking for a transparent and adaptable numerical scheme.

\section{Conclusions}\label{Conclusiones}

This work presented a step-by-step implementation of the RK4 method to numerically solve the Lindblad master equation, applied to the case of a single coherent field mode subject to decay. The approach emphasizes clarity and pedagogical accessibility, providing an explicit numerical routine suitable for students and researchers beginning in the study of open quantum systems.

The methodology proved capable of reproducing not only the expected transient dynamics of the average photon number, but also its steady-state behavior, confirming the internal consistency and physical precision of the numerical scheme. The explicit recovery of the steady state from the Liouvillian spectrum also offers a meaningful tool for future studies involving thermalization, equilibrium, or non-Hermitian effects.

By modifying only the Hamiltonian and collapse operators, the same framework can be applied to a wide variety of dissipative systems, including multimode fields, atom–field interactions, or thermal reservoirs.

Overall, this implementation provides a solid foundation for future explorations, combining physical transparency, computational control, and educational value within a reproducible open source environment. It is particularly useful in contexts where understanding the underlying structure of the evolution equation is essential, offering an open alternative to high-level packages that obscure the numerical treatment of dissipation.

\section*{Acknowledgments}
L. Hernández-Sánchez acknowledges the Instituto Nacional de Astrofísica, Óptica y Electrónica (INAOE) for the collaboration scholarship granted and the Consejo Nacional de Humanidades, Ciencias y Tecnologías (CONAHCYT) for the SNI III assistantship scholarship (No. CVU: 736710). In addition, the work of I.A. Bocanegra-Garay is supported by Spanish MCIN with funding from the European Union Next Generation EU (PRTRC17.I1) and the Consejeria de Educacion from Junta de Castilla y Leon through the QCAYLE project, as well as the grant number PID2023-148409NB-I00 MTM funded by AEI/10.13039/501100011033, and RED2022-134301-T. The financial support of the Department of Education, the Junta de Castilla y Leon, and the FEDER Funds is also gratefully acknowledged (Reference: CLU-2023-1-05).

\bibliographystyle{unsrt} 
%


\end{document}